\documentclass[11pt,a4paper]{article}
\usepackage[hyperref]{main}
\usepackage{times}
\usepackage{latexsym}
\usepackage{graphicx}
\usepackage{caption}
\usepackage{amsmath}
\usepackage{amssymb}
\usepackage{amsfonts}
\usepackage{booktabs}
\usepackage[bottom]{footmisc}
\usepackage{url}
\usepackage[linesnumbered,lined,algoruled]{algorithm2e}
\usepackage{multicol}
\usepackage{qtree}
\graphicspath{{png/}}

\aclfinalcopy 

\title{Efficient Parallelization of a Ubiquitous Sequential Computation}

\author{Franz A. Heinsen \\
	{\tt franz@glassroom.com} \\ }

\date{October 27, 2023}

\newcommand{\bigO}{\mathcal{O}}
\DeclareMathOperator{\lcse}{LCSE}
\DeclareMathOperator{\tail}{tail}
\DeclareMathOperator{\cat}{cat}

\newcommand{\cum}[1]{\overset{\text{\textsmaller{cum}}}{#1}}

\newcommand{\astar}{a^*}
\newcommand{\bstar}{b^*}

\SetCommentSty{commentstyle}

\begin{document}
\maketitle

\begin{abstract}
We find a succinct expression for computing the sequence $x_t = a_t x_{t-1} + b_t$ in parallel with two prefix sums, given $t = (1, 2, \dots, n)$, $a_t \in \mathbb{R}^n$, $b_t \in \mathbb{R}^n$, and initial value $x_0 \in \mathbb{R}$. On $n$ parallel processors, the computation of $n$ elements incurs $\bigO(\log n)$ time and $\bigO(n)$ space. Sequences of this form are ubiquitous in science and engineering, making efficient parallelization useful for a vast number of applications. We implement our expression in software, test it on parallel hardware, and verify that it executes faster than sequential computation by a factor of $\frac{n}{\log n}$.\footnote{Source code for replicating our results is available online at \href{https://github.com/glassroom/heinsen_sequence}{https://github.com/glassroom/heinsen\_sequence}.}
\end{abstract}

\section{Summary}\label{sec:summary}

Sequences of the form $x_t = a_t x_{t-1} + b_t$ are ubiquitous in science and engineering. For example, in the natural sciences, such sequences can model quantities or populations that decay or grow by a varying rate $a_t > 0$ between net inflows or outflows $b_t$ at each time step $t$. In economics, such sequences can model investments that earn a different rate of return $a_t = (1 + r_t)$ between net deposits or withdrawals $b_t$ over each time period $t$. In engineering applications, such sequences are often low-level components of larger models, {\em e.g.}, linearized recurrent neural networks whose layers decay token features in a sequence of tokens.

Given a finite sequence $x_t = a_t x_{t-1} + b_t$ with $n$ steps, $t = (1, 2, \dots, n)$, where $a_t \in \mathbb{R}^n$, $b_t \in \mathbb{R}^n$, and initial value $x_0 \in \mathbb{R}$, it's not immediately obvious how one would compute all elements in parallel, because each element is a non-associative transformation of the previous one. {\em In practice, we routinely see software code that computes sequences of this form one element at a time.}

The vector $\log x_t$ is computable as a composition of two cumulative, or prefix, sums, each of which {\em is} parallelizable:

\begin{equation}\label{eq:parallel_log_sequence}
	\log x_t = \astar_t + \log \left( x_0 + \bstar_t \right) \\
\end{equation}

where $\astar_t$ and $\bstar_t$ are the two prefix sums:

\begin{equation}\label{eq:astar_and_bstar}
	\begin{aligned}
		\astar_t & = \cum{\sum\limits_t} \log a_t \\
		\bstar_t & = \cum{\sum\limits_t} e^{\log b_t - \astar_t}. \\
	\end{aligned}
\end{equation}

The operator $\cum\sum$ computes a {\em vector} whose elements are a prefix sum, {\em i.e.}, a cumulative sum.

We obtain $x_t$ with elementwise exponentiation:

\begin{equation}\label{eq:parallel_sequence}
	x_t = e^{\astar_t + \log (x_0 + \bstar_t)}. \\
\end{equation}

Prefix sums are associative,\footnote{
	Given a sequence $a, b, c$, 
	$$
	\cum\sum \Big( a, \cum\sum \Big( b, c \Big) \Big) 
	=
	\cum\sum \Big( \cum\sum \Big( a, b \Big), c \Big).
	$$
} making it possible to compute them by parts in parallel. Well-known parallel algorithms for efficiently computing the prefix sum of a sequence with $n$ elements incur $\bigO(\log n)$ time and $\bigO(n)$ space on $n$ parallel processors \cite{10.1145/322217.322232} \cite{10.1145/7902.7903}. Prefix sums generalize to any binary operation that is associative, making them a useful primitive for many applications \cite{BlellochTR90} and data-parallel models of computation \cite{10.5555/91254}. Many software frameworks for numerical computing provide efficient parallel implementations of the prefix sum.

The computation of two prefix sums has the same computational complexity on $n$ parallel processors as a single prefix sum: $\bigO(\log n)$ time and $\bigO(n)$ space. The computation of $n$ elementwise operations ({\em e.g.}, logarithms and exponentials) on $n$ parallel processors incurs constant time and, if the computation is {\em in situ}, no additional space.

If any $a_t < 0$, any $b_t < 0$, or $x_0 < 0$, one or more of the logarithms computed in the interim will be in $\mathbb{C}$, but all elements of $x_t$ will always be in $\mathbb{R}$, because they are defined as multiplications and additions of previous elements in $\mathbb{R}$, which is closed under both operations.

\section{Compared to Blelloch's Formulation}\label{sec:compared_to_blelloch}

Blelloch's formulation for computing first-order linear recurrences as a composition of prefix sums \cite{BlellochTR90} is more general, expressed in terms of a binary operator $\oplus$ that is associative and a second binary operator $\otimes$ that either is associative or can be transformed into an associative one via the application of a third binary operator.

Our formulation applies only to the most common case, real numbers, with scalar sum and multiplication as the first and second operators, making each step non-associative. We find a succinct, numerically stable expression that is readily implementable with widely available, highly-optimized implementations of the prefix sum.

\section{Proof}\label{sec:proof}

We are computing $x_t = a_t x_{t-1} + b_t$, for $t = (1, 2, \dots, n)$, with $a_t \in \mathbb{R}^n$, $b_t \in \mathbb{R}^n$, and initial value $x_0 \in \mathbb{R}$. Expand the expression that computes each element, $x_1, x_2, \dots, x_n$, to make it a function of $x_0$ and all trailing elements of $a_t$ and $b_t$, and factor out all trailing coefficients:

\begin{equation}\label{eq:element_by_element_expressions}
	\thinmuskip=0.5mu\medmuskip=0.5mu\thickmuskip=1mu
	\begin{aligned}
		x_1
		& = a_1 x_0 + b_1 \\
		& = a_1 \left( x_0 + \frac{b_1}{a_1} \right) \\
		\\
		x_2
		& = a_2 x_1 + b_2 \\
		& = a_1 a_2 \left( x_0 + \frac{b_1}{a_1} + \frac{b_2}{a_1 a_2} \right) \\
		& \vdots \\
		x_n
		& = a_n x_{n-1} + b_n \\
		& = \Big( \! \prod\limits_t a_t \! \Big) \bigg( \! x_0 + \frac{b_1}{a_1} + \frac{b_2}{a_1 a_2} + \dots + \frac{b_n}{{\scriptstyle \prod\limits_t} a_t} \! \bigg). \\
	\end{aligned}
\end{equation}

Combine all expressions in \eqref{eq:element_by_element_expressions} into one expression that computes all elements of vector $x_t$:

\begin{equation}
	\thinmuskip=0.5mu\medmuskip=1mu\thickmuskip=2mu
	\begin{aligned}
		x_t
		& = \bigg( \cum{\prod\limits_t} a_t \bigg) \odot \Bigg( x_0 + \cum{\sum\limits_t} \frac{b_t}{\cum{\scriptstyle \prod\limits_t} a_t} \Bigg) \\
		& = \bigg( \cum{\prod\limits_t} a_t \bigg) \odot \Bigg( x_0 + \cum{\sum\limits_t} \exp \Bigg( \log \frac{b_t}{\cum{\scriptstyle \prod\limits_t} a_t} \Bigg) \!\! \Bigg) \\
		& = \bigg( \cum{\prod\limits_t} a_t \bigg) \odot \Bigg( x_0 + \cum{\sum\limits_t} e^{\log b_t - \cum\sum_t \log a_t } \Bigg), \\
	\end{aligned}
\end{equation}

where the operators $\cum\prod$ and $\cum\sum$ compute {\em vectors} whose elements are, respectively, a cumulative product and sum, and $\odot$ denotes an elementwise or Hadamard product.

Taking the logarithm on both sides, we obtain:

\begin{equation}\label{eq:derivation_of_astar_and_bstar}
	\small
	\thinmuskip=1mu\medmuskip=1mu\thickmuskip=1mu
	\log x_t
	=
	\underbrace{ \cum{\sum\limits_t} \log a_t }_{\textstyle \astar_t}
	+
	\log \Bigg(
		x_0 + \underbrace{ \cum{\sum\limits_t} e^{
				\log b_t - \overbrace{\scriptstyle \cum\sum_t \log a_t }^{\textstyle \astar_t}
			} }_{\textstyle \bstar_t}
	\Bigg), \\
\end{equation}

which is the same as \eqref{eq:parallel_log_sequence}.

\section{Implementation}\label{sec:implementation}

\begin{figure}[t]
	\vskip 0.1in
	\begin{center}
		\centerline{\includegraphics{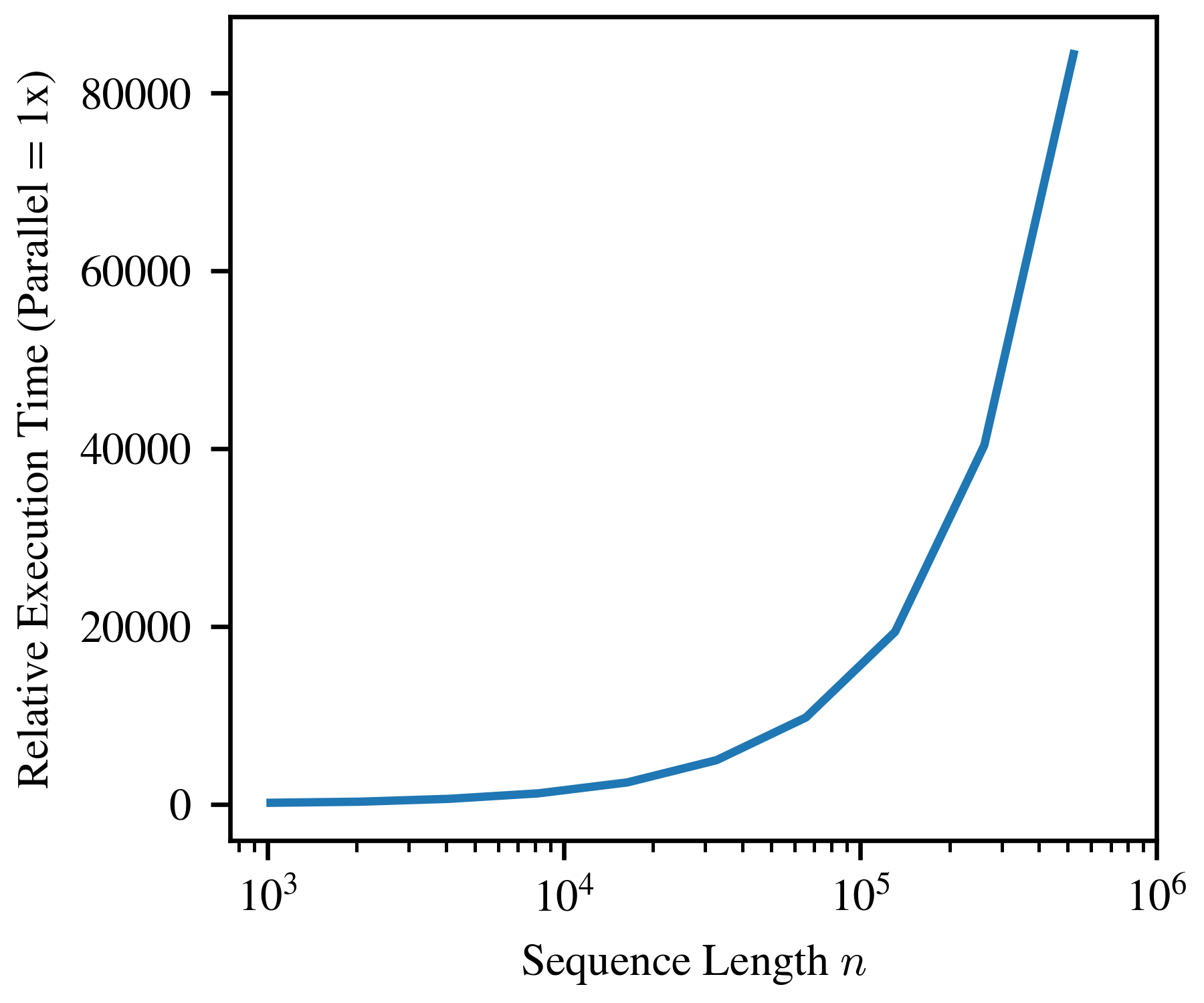}}
		\caption{Time to compute $n$ elements sequentially, relative to parallel computation, on an Nvidia GPU. Each point is the mean of 30 runs.}
		\label{fig:relative_execution_time}
	\end{center}
	\vskip -0.2in
\end{figure}

We implement \eqref{eq:parallel_sequence} in software. For numerical stability and slightly improved efficiency, we modify the computation of $x_t$ as follows:

\begin{equation}
	x_t = e^{\astar_t + \tail(\lcse(\cat( \log x_0, \log b_t - \astar_t )))},
\end{equation}

where $\cat(\cdot)$ denotes concatenation, $\tail(\cdot)$ removes its argument's first element, and

\begin{equation}
	\lcse(\cdot) := \log\cum\sum\exp(\cdot),
\end{equation}

commonly provided as the ``LogCumSumExp'' function by software frameworks for numerical computing, applying the familiar log-sum-exp trick as necessary for numerical stability, and delegating parallel computation of the internal prefix sum to a highly-optimized implementation. 

We test our implementation on parallel hardware and verify that it executes faster than sequential computation by a factor of $\frac{n}{\log n}$ (Figure \ref{fig:relative_execution_time}).

\bibliography{main}
\bibliographystyle{main}

\end{document}